\documentclass[11pt]{article}
\usepackage[unicode=true]{hyperref}
\usepackage{amsmath}
\usepackage{cite}
\usepackage{amsfonts}
\usepackage{amssymb}
\usepackage{authblk}
\usepackage[pdftex]{color,graphicx}
\newtheorem{Theorem}{Theorem}

\newtheorem{Cor}{Corollary}

\newtheorem{Lemma}{Lemma}

\newcommand{\Tr}{\text{Tr}}
\newcommand{\hH}{\text{Herm}}
\newcommand{\ZZ}{\mathbb{Z}}
\newcommand{\RR}{\mathbb{R}}
\newcommand{\CC}{\mathbb{C}}
\newcommand{\cC}{\mathcal{C}}
\newcommand{\iI}{\mathcal{I}}
\newcommand{\vV}{\mathcal{V}}
\newcommand{\wW}{\mathcal{W}}

\newcommand{\sS}{\mathcal{S}}
\newcommand{\oO}{\mathcal{O}}
\newcommand{\Cl}{\text{Cl}}  
\newcommand{\Span}[1]{\ensuremath{ \langle #1 \rangle }}
\newcommand{\Sp}{\text{Sp}}
\newcommand{\cnc}{\text{cnc}}

\newcommand{\set}[1]{\ensuremath{ \lbrace #1 \rbrace }}

\definecolor{dur}{cmyk}{0,1,1,0.3}

\newcommand{\id}{\mathbb{1}}

\usepackage{enumitem}
\usepackage{soul}

\usepackage{bbold}
\usepackage{tikz}
\usetikzlibrary{decorations.pathreplacing,angles,quotes}

\usetikzlibrary{matrix,arrows,decorations.pathmorphing}
\usepackage{tikz-cd}
\usetikzlibrary{arrows}

\title{ On the extremal points of the $\Lambda$-polytopes and classical simulation of quantum computation with magic states}
 
\author[1]{Cihan Okay}
\author[2,3]{Michael Zurel}
\author[2,3]{Robert Raussendorf}
\affil[1]{Department of Mathematics, Bilkent University, Ankara, Turkey}
\affil[2]{Department of Physics and Astronomy, University of British Columbia, Vancouver, BC, Canada}
\affil[3]{Stewart Blusson Quantum Matter Institute, University of British Columbia, Vancouver, BC, Canada} 
\date{\today}                     
\begin{document}
\maketitle
\begin{abstract}
We investigate the $\Lambda$-polytopes, a convex-linear structure recently defined and applied to the classical simulation of quantum computation with magic states by sampling. There is one such polytope, $\Lambda_n$, for every number $n$ of qubits. We establish two properties of the family $\{\Lambda_n, n\in \mathbb{N}\}$, namely (i) Any extremal point (vertex) $A_\alpha \in \Lambda_m$ can be used to construct vertices in $\Lambda_n$, for all $n>m$. (ii) For vertices obtained through this mapping, the classical simulation of quantum computation with magic states can be  efficiently reduced to the classical simulation based on the  preimage $A_\alpha$.  In addition, we describe a new class of vertices in $\Lambda_2$ which is outside the known classification. While the hardness of classical simulation remains an open problem for most extremal points of $\Lambda_n$,  the above results extend efficient classical simulation of quantum computations beyond the presently known range. 
\end{abstract}

\section{Introduction}  

The question of precisely which property of quantum mechanics is responsible for the speedup in quantum computation to date remains open. Various candidates have been put forward, such as superposition and interference \cite{sup}, entanglement \cite{Vidal1, VdN}, and largeness of Hilbert space \cite{Feyn}, but there is always a counterexample that stops broad generalization; see e.g. \cite{Stab,VdN2,TETBU, Illu}.
To identify new candidates, or perhaps even refine the question, one may study classical simulation algorithms for quantum systems, and specifically how the efficiency of such algorithms breaks down when pushed towards the regime of universal quantum computation. 

 The classical simulation method of present interest is based on sampling from the $\Lambda$-polytope \cite{HVM}. Before turning to it, to provide context,  we briefly  discuss other classical simulation methods for quantum computation, and the general structure that unifies them.

Classical simulation techniques for quantum computation typically seek to exploit proximity to a simple,  ``effectively classical'' reference point.  The characteristic simplifying feature of such reference points varies; examples are locality \cite{loc},  stabilizerness \cite{Stab}, matchgates \cite{MG,JW2},  and the positivity of Wigner functions \cite{NegWi,M1}. Furthermore, often a natural measure exists for the distance  between the setting of interest  and the classical reference point. This can be,  in case of locality as the simplifying feature, the bond dimension in MPS states \cite{Vidal1}, or the graph-theoretic measure of rankwidth in classical simulation of measurement based quantum computation on graph states \cite{VdN}.  If stabilizer-ness is the simplifying feature,  measures of state magic \cite{NegWi, Hakop, M1,M1b,M2,M3,M4,M5}  such as Wigner function negativity or stabilizer rank  quantify the hardness of classical simulation. 

Irrespective of the concrete approach, these classical simulation methods share four general features: (i) A sector in which the classical simulation is efficient, (ii) a physical property that characterizes this sector, (iii) a physical property of universal quantum computation that obstructs efficiency of the simulation method in the general case, and (iv) a distance measure from the setting of interest to its closest efficiently simulable setting,  governing the hardness of classical simulation.   

In this paper, we are concerned with the recently introduced method of simulating quantum computation based on sampling from the $\Lambda$-polytope \cite{HVM}. Being a sampling algorithm, it is closely related to the sampling algorithms invoking Wigner functions and other  quasiprobability distributions \cite{NegWi, Hakop, Galv1, Galv2, ReWi, raussendorf2020phase}. However, there is also an important difference: in the present method, no negativity ever occurs!

This prompts the question of where quantumness is hiding in this scenario. If quantumness cannot be attributed to negativity, then to what else?  To summarize the state of knowledge about the classical simulation by sampling from the $\Lambda$-polytope \cite{HVM}, of the  above general characteristics (i) - (iv), we presently only have a glimpse of (i). Namely, it is known that in the   subpolytope spanned by the so-called cnc vertices, classical simulation is efficient \cite{raussendorf2020phase}; and this includes the domain of the stabilizer formalism \cite{Stab} as a strict subset. But it is presently unknown how far the efficiently simulable region extends, which physical property this efficiency is to be attributed to, and which opposing physical property would render the simulation of the general case inefficient.  

This paper is a first approach to the systematic study of the $\Lambda$-polytopes. We investigate the structure of the extremal points of the polytopes $\Lambda_n$, asking: {\em{Can the vertices of $\Lambda_n$ be constructed from smaller parts? And if so, can the classical simulation of  quantum computation  based on such vertices be reduced to the classical simulation of those parts?}}
---Here, we give affirmative first answers to these questions, extending the classically efficiently simulable sector of the model \cite{HVM}.

The main results of this paper are:
\begin{itemize}
\item{Theorem~\ref{thm:Phi-map} (Section~\ref{mapping}):  For every vertex $X\in  \Lambda_m$, every projector onto an $(n-m)$-qubit stabilizer state $|\sigma\rangle$, $\Pi_\sigma=|\sigma\rangle \langle \sigma|$, and every  $n$-qubit Clifford unitary $U$, it holds that $U(X\otimes \Pi_\sigma)U^\dagger$ is a vertex of $\Lambda_{n}$.}

\item{Theorem~\ref{T1} (Section~\ref{sim}): The classical simulation of quantum computation with magic states (QCM), with the  $n$-qubit initial state $\tilde{X} = U(X\otimes \Pi_\sigma)U^\dagger$, can be efficiently reduced to  the classical simulation of QCM with the  $m$-qubit initial state $X$.}
 
\item{Theorem~\ref{thm:update-rules} (Section~\ref{BeyondCNC}): We describe a new family of vertices $X\in \Lambda_2$, i.e., of two qubits, which is outside the closed and  noncontextual (cnc) classification \cite{raussendorf2020phase}.   It is the first explicitly known family, including the update under Pauli measurements, beyond cnc.}

\end{itemize}
The significance of the first result is that whenever we learn about a new vertex of  $\Lambda_m$, it implies an infinite family of vertices on $n> m$ qubits.  The second result means that, for the entire infinite family of vertices resulting from $X$, the hardness of classical simulation of QCM is essentially given by the hardness of simulating QCM on $X$.  This result,  together with the new class of efficiently updateable vertices provided by the third result, extends the efficiently classically simulable sector for the present method.
\medskip

This paper is organized as follows. In Section~\ref{BG} we review background material. We state the definition of the state polytopes $\Lambda_n$ and provide a motivation for studying them. In Section~\ref{mapping} we show how to construct vertices of the state polytope  $\Lambda_n$ from vertices of the polytope  $\Lambda_m$, for  $n>m$. Section~\ref{sim} is concerned with the efficiency of classical simulation of quantum computation with magic states.  For the cases where a vertex under consideration is the result of  the map defined by Theorem~\ref{thm:Phi-map}, we describe a reduction to the classical simulation of the preimage of  the map.  In Section~\ref{BeyondCNC}, we describe a new family of two-qubit vertices that is outside the presently known cnc classification. We conclude in Section~\ref{Concl}.

\section{Background}\label{BG}

 This section begins with a short summary of quantum computation with magic states (Section~\ref{magi}), which provides our main motivation for studying the $\Lambda$-polytopes. We then state the definition of the state polytopes $\Lambda_n$ and summarize the properties presently known of them (Section~\ref{BasDef}). We conclude with a discussion of the role of the tensor product for the present construction (Section~\ref{TP}).

\subsection{Quantum computation with magic states}\label{magi}

 The $\Lambda$-polytopes have found a surprising application in describing universal quantum computation with magic states in terms of sampling \cite{HVM}. Therefore, before we discuss the $\Lambda$-polytopes themselves we provide a brief summary of that scheme of quantum computation.

Quantum computation with magic states (QCM) \cite{GT,NC,BK} is a scheme for universal quantum computation, closely related to the circuit model. There is also an important difference. The unitary gates in QCM are not universal. This is compensated by the provision of so-called magic states which restore computational universality. In this way, computational power shifts from the gates to the initial states. As an aside, from a practical point of view QCM is very advantageous for fault-tolerant quantum computation \cite{BK}.

There are two types of operations in QCM, the ``free'' operations and the ``resources''. The free operations are (i) preparation of all stabilizer states, (ii) the Clifford unitaries, and (iii) measurement of all Pauli observables.

The resource are arbitrarily many copies of the state
\begin{equation}\label{MagStat}
|T\rangle  = \frac{|0\rangle + e^{i\pi/4}|1\rangle}{\sqrt{2}}.
\end{equation}
The state $|T\rangle $ is called a ``magic state'', as it restores computational universality given the other above operations.

The distinction between free operations and resources in QCM is motivated by the Gottesman-Knill theorem which says that  free operations alone can be efficiently classically simulated. 

\subsection{The $\Lambda$-polytopes} \label{BasDef}

\subsubsection{Definition}

Let  $\hH(2^n)$  denote the set of Hermitian operators on the $n$-qubit Hilbert space ${\cal H}=\CC^{2^n}$.
We recall the definition of the polytope $\Lambda_n$ from \cite{HVM,Heim}:
\begin{equation}\label{eq:poly}
\Lambda_n = \set{X\in \hH_1(2^n)|\; \Tr(X|\sigma\rangle \langle \sigma|)\geq 0,\;\forall |\sigma\rangle \in \sS_n }
\end{equation}
where $\hH_1(2^n) \subset \hH(2^n)$ denotes the subset of Hermitian matrices  of trace $1$ and $\sS_n$ denotes the set of pure $n$-qubit stabilizer  states. The set of vertices of $\Lambda_n$ is denoted by $\set{A_\alpha|\;\alpha\in\vV_n}$.  We note that in \cite{Heim} the polytope $\Lambda_n$ is  used as a convenient tool to study the facets of the stabilizer polytope.

The state spaces $\Lambda_n$ are thus generalizations of the sets of proper density matrices. Namely, for density matrices we require that  with respect to any quantum state, the Born rule yields a probability for finding that state in measurement, i.e., a  nonnegative real number. For the elements of $\Lambda_n$, we relax this condition. Namely, we only require  it for stabilizer states. \medskip

 In what follows, we will often expand vertex operators in the basis of $n$-qubit Pauli operators. Writing $E_n$ for the vector space $(\ZZ_2)^{n}\times (\ZZ_2)^n$,  the Pauli operators are
$$
T_v = i^{v_Z\cdot v_X} Z(v_Z)X(v_X), \;\; \forall (v_Z,v_X)=v \in E_n,
$$ 
where $v_Z\cdot v_X$ is computed mod $2$, and $X(v_X):=\bigotimes_{i=1}^n (X_i)^{[v_X]_i}$,  $Z(v_Z):=\bigotimes_{i=1}^n (Z_i)^{[v_Z]_i}$. Sometimes we will write $T_v^{(n)}$ to emphasize that the operator is an $n$-qubit Pauli operator.

$E_n$ comes with a symplectic form defined by 
$$[v,w]=v_Z^T w_X + v_X^T w_Z$$
for all $v=(v_Z,v_X),w=(w_Z,w_X)\in E_n$.
We denote the canonical symplectic basis of $E_n$ by   $\set{x_1,\cdots,x_n,z_1,\cdots,z_n}$.  Also it is convenient to introduce the notation $y_n=x_n+z_n$. We write $\Sp_{2n}(\ZZ_2)$ for the group of symplectic transformations   acting on $E_n$.  For a subspace $W\subset E_n$ let $W^\perp$ denote  the subspace $\set{v\in E_n|\;[v,w]=0,\;\forall w\in W}$. A subspace $J\subset E_n$ is called isotropic if $J\subset J^\perp$.  The collection of maximal isotropic subspaces of $E_n$ is denoted by $\iI(E_n)$.

\subsubsection{Properties of  $\Lambda_n$}

Beyond the defining relations and the application to QCM \cite{HVM}, to date  two general structural facts are known about the family $\{\Lambda_n,\,n\in \mathbb{N}\}$ of state spaces. They are
\begin{enumerate}
 \item{For all $n$,  $\Lambda_n$ is a polytope, hence  it is bounded and the set of its extremal points is finite.}
 \item{\label{PointN} For all $n$, $\Lambda_n$ contains all $n$-qubit density matrices, and it maps into itself under conjugation by Clifford unitaries and Pauli measurements.  }
\end{enumerate}

\paragraph{Absence of negativity.}  As a consequence of the above point \ref{PointN}, negativity of quasiprobability distributions can be entirely dispensed with in the description of QCM. There exists a representation \cite{HVM} for which all quantum states, as well as all quantum operations necessary for QCM, can be represented by probabilities and conditional probabilities, respectively, giving rise to a hidden variable model (HVM) describing universal quantum computation by sampling.

Namely, we have the following result invoking the state polytopes $\Lambda_n$ of Eq.~(\ref{eq:poly}).
\begin{Theorem}[\cite{HVM}]\label{MT}
For all numbers of qubits $n\in \mathbb{N}$, (i) each $n$-qubit quantum state $\rho$ can be represented by a probability function $p_\rho: \mathcal{V}_n \to \mathbb{R}_{\ge0}$, $\rho = \sum_{\alpha\in\mathcal{V}_n} p_\rho(\alpha)\, A_\alpha$.

(ii) For the state update under Pauli measurements it holds that
$\Pi_{a,s} A_\alpha \Pi_{a,s} = \sum_{\beta \in \mathcal{V}_n} q_{\alpha,a}(\beta,s)\, A_\beta$.
For all $a \in E_n$, $\alpha\in \mathcal{V}_n$, the  $q_{\alpha,a}: \mathcal{V}_n\times \mathbb{Z}_2\to \mathbb{R}_{\ge0}$ are probability functions, 

(iii) Denote by $P_{\rho,a}(s)$ the probability of obtaining outcome $s$ for a measurement of $T_a$ on the state $\rho$. Then,
the Born rule $P_{\rho,a}(s)=\text{{\em{Tr}}}(\Pi_{a,s}\rho)$ takes the form $\text{{\em{Tr}}}(\Pi_{a,s}\rho) = \sum_{\alpha \in {\cal{V}}_n}p_\rho(\alpha)Q_a(s|\alpha)$.
Therein, $Q_a(s|\alpha)$ is given by  $Q_a(s|\alpha) :=\sum_{\beta \in \mathcal{V}_n} q_{\alpha,a}(\beta,s)$. Hence $0\leq Q_a(s|\alpha) \leq 1$, for all $a,s,\alpha$.
\end{Theorem}
The theorem does not speak about the Clifford gates; however, those are also positively represented (see the SM of \cite{HVM}, Sec.~I). Alternatively, the Clifford gates may be propagated past the last measurement in any given  QCM circuit, and then discarded. They do not add computational power beyond that afforded by the Pauli measurements~\cite{BravyiSmolin2016}.

The above Theorem~\ref{MT}, and its application to the classical simulation of quantum computation with magic states, is our main motivation for studying the polytopes $\Lambda_n$.

\paragraph{Cnc vertices.}  There is an infinite family of vertices of the $\Lambda$-polytopes, the so-called  cnc vertices, which is completely understood. Any cnc
 vertex in $\vV_n$ has the form
\begin{equation}\label{eq:cnc-type}
A^\gamma_\Omega = \frac{1}{2^n} \sum_{v\in \Omega} (-1)^{\gamma(v)} T_v
\end{equation}
where $\Omega \subset E_n$ is a maximal cnc set, to be defined below. The set of $n$-qubit cnc-type vertices are denoted by $\vV_n^\cnc$.

``Cnc'' in `cnc set $\Omega$' stands for ``closed  noncontextual'', and means the following. (Closure) A set $\Omega \subset E_n$ is closed if $v,w\in \Omega$ and $[v,w]=0$ implies that $v+w \in \Omega$. (Noncontextuality) $\Omega$ admits a consistent  noncontextual value assignment $\gamma: \Omega \longrightarrow \mathbb{Z}_2$. That is, the eigenvalue found in the measurement of the Pauli observable $T_v$ is $(-1)^{\gamma(v)}$. The consistency requirement is that for a triple of commuting Pauli observables $T_v$, $T_w$, $T_{v+w}$, with $T_{v+w}= (-1)^{\beta(v,w)}T_vT_w$, it holds that $\gamma(v+w)=\gamma(v)+\gamma(w)+\beta(v,w) \mod 2$.

The reason for imposing the consistency condition is that if the three observables $T_v$, $T_w$, $T_{v+w}$ are simultaneously measured, on any quantum state, then the measurement outcomes  of these observables satisfy the above constraint with certainty. Hence we require the same of all considered value assignments $\gamma$. The consistency condition also explains the need for the closure property. If $v,w\in \Omega$ then the expectation values for $T_v$, $T_w$ are extremal. The expectation value for $T_{v+w}$ must then also be extremal, hence we require $v+w \in \Omega$ for $A_\Omega^\gamma$ to reproduce this.

The following is known about the cnc vertices:
\begin{enumerate}
\item{For any $n\in \mathbb{N}$, the convex hull of $\vV_n^\cnc$  is closed under Pauli measurement. That is, for any projector $\Pi_{a,s}$ onto the eigenspace of $T_a$ corresponding to the eigenvalue $(-1)^s$ it holds that if $\Tr(\Pi_{a,s}A_\Omega^\gamma)>0$ then $\Pi_{a,s} A_\Omega^\gamma \Pi_{a,s}$ is a probabilistic linear combination of  cnc vertices (Theorem~2 of \cite{raussendorf2020phase}.)}
\item{The update of the pair $(\Omega,\gamma)$ under Pauli measurement is computationally efficient (Theorem~3 of \cite{raussendorf2020phase}).}
\item{\label{PointPRA}The phase point operators $A_\Omega^\gamma$ are extreme points of the respective polytopes $\Lambda_n$~\cite{HVM,Heim}.}
\item{The cnc sets $\Omega$ are classified for all $n\in \mathbb{N}$ (Theorem~1 of \cite{raussendorf2020phase}; also see Theorem~2 of \cite{kirby2019contextuality}).} 
\end{enumerate}
Item~\ref{PointPRA} provides an element of continuity between the simulation method \cite{HVM} of present interest and the earlier simulation schemes based on Wigner functions \cite{NegWi,ReWi,raussendorf2020phase}.  Namely, the phase point operators $A_\Omega^\gamma$ were initially devised for a classical simulation scheme \cite{raussendorf2020phase} of quantum computation with magic states in which negativity can arise. The latter is the multiqubit counterpart to the initial work on the role of Wigner function negativity for quantum computation \cite{NegWi}, which applies only to odd Hilbert space dimension. 

\subsection{Status of the tensor product}\label{TP}

Given two extremal points $A_\alpha \in \Lambda_n$ and $A_\beta\in \Lambda_m$, is the tensor product $A_\alpha \otimes A_\beta$ an extremal point of $\Lambda_{n+m}$?---This is not generally the case; in fact, $A_\alpha \otimes A_\beta$ may not even be contained in $\Lambda_n$. We can show this by example. Consider the one-qubit vertex \cite{HVM}
$$
A_{0} = \frac{I+X+Y+Z}{2}.
$$
Then it holds that $A_{0} \otimes A_{0} \not \in \Lambda_2$. To see this, consider the stabilizer (Bell) state $|B_{11}\rangle$ defined by the stabilizer relations $Z_1\otimes Z_2|B_{11}\rangle =-|B_{11}\rangle$, $X_1\otimes X_2|B_{11}\rangle =-|B_{11}\rangle$.  With these relations, $\langle B_{11}| A_{0} \otimes A_{0}  |B_{11}\rangle = -1/2$. Hence, with Eq.~\ref{eq:poly}, $A_{0} \otimes A_{0} \not \in \Lambda_2$. In the terminology of the Pusey-Barrett-Rudolph (PBR) theorem \cite{PBR}, the HVM of Theorem~\ref{MT} does not satisfy the condition of ``preparation independence'', thereby evading the consequences of the PBR theorem.

From the above discussion it appears that the tensor product does not play a particular role in the HVM described by Theorem~\ref{MT}. Or does  it?

\section{Mapping vertices of $\Lambda_{m}$ to $\Lambda_n$}\label{mapping}

In the introduction we asked whether the extremal points of the state polytopes $\Lambda_n$ can be constructed from smaller parts, and Section~\ref{TP} concluded with the question of whether there is any significant role for the tensor product in the present formalism. It turns out that the two questions are related through a common answer.

 In this section we construct a map $\Lambda_m\to \Lambda_n$, for any $m<n$, that sends a vertex of $\Lambda_m$ to a vertex of $\Lambda_n$. Moreover, under this map $\vV_{m}$  maps injectively into $\vV_{n}$.
For the statement of our result we recall the definition of stabilizer projectors. Let $\Pi_{J,s}$ denote the projector corresponding to the pair $(J,s)$ consisting of an isotropic subspace $J\subset E_n$ and a value assignment $s:J\to \ZZ_2$.  
It has the form
$$
\Pi_{J,s} = \frac{1}{|J|} \sum_{v\in J} (-1)^{s(v)} T_v
$$
where $|\,|$ denotes the number of elements. For each stabilizer state $|\sigma \rangle \in \sS_n$ the projector $\Pi_\sigma=|\sigma\rangle \langle \sigma|$ can be uniquely written as a projector of the form $\Pi_{I,s}$ for some maximal isotropic subspace $I\subset E_n$ and a value assignment $s:I\to \ZZ_2$.

\begin{Theorem}\label{thm:Phi-map}
Let  $J\subset E_n$ be an isotropic subspace of dimension  $d=n-m$ and $S\in \Sp_{2n}(\ZZ_2)$ such that $S(J_0)=J$ where $J_0=\Span{x_{m+1},\cdots,x_n}$. 
Given  value assignments $r_0:J_0\to \ZZ_2$ and $r:J\to \ZZ_2$ define a linear map
$$
\Phi_{J,r,S|_{E_{m}}}: \hH(2^{m}) \to \hH(2^n)
$$
by the formula
\begin{equation}\label{eq:Phi-map}
\Phi_{J,r,S|_{E_{m}}}(X) = U( X\otimes \Pi_{J_0,r_0} )U^\dagger
\end{equation}
where $U$ is a Clifford unitary that implements $S$ and satisfies   $U\Pi_{J_0,r_0} U^\dagger = \Pi_{J,r}$. Then the map $\Phi_{J,r}=\Phi_{J,r,S|_{E_{m}}}$ satisfies the following properties:
\begin{enumerate}
\item   The image of $\Phi_{J,r}$ is given by 
 $\set{(\Pi_{J,r} Y \Pi_{J,r})/Tr(Y\Pi_{J,r})|\;Y\in \Lambda_n \text{ and } \Tr(Y\Pi_{J,r})\neq 0 }$.
\item $\Phi_{J,r}$ is injective and maps a vertex  $X\in \Lambda_{m}$ to a vertex of $\Lambda_n$.
\item A vertex $X$ is of cnc-type (see Eq.~(\ref{eq:cnc-type}))  if and only if $\Phi_{J,r}(X)$ is of cnc-type.
\end{enumerate}
\end{Theorem}
 Presently, very little is known about the vertices of the polytopes $\Lambda_n$, for $n\geq 3$; and Theorem~\ref{thm:Phi-map} extends that knowledge base. Essentially, the only fact already known is that the  cnc construction \cite{raussendorf2020phase} provides vertices of $\Lambda_n$ for all $n$. Cnc vertices have the property that they consist of a core tensored  with a stabilizer tail, cf. Lemma~11 in \cite{raussendorf2020phase}. Therein, the cores are related to Majorana fermions and the Jordan-Wigner transformation. Now, Theorem~\ref{thm:Phi-map} establishes that {\em{all}} vertices of $\Lambda_n$, for any $n$, have the same overall  structure: they consist of a core tensored with a stabilizer tail.\medskip 

We now prepare for the proof of Theorem~\ref{thm:Phi-map}. We first consider the case $(J,r)=(J_0,r_0)$ so that $S$ is the identity matrix, and also the Clifford unitary $U$ is chosen to be the identity matrix.
For $X\in \hH(2^{m})$ we can write
$$
X = \frac{1}{2^{m}} \sum_{v\in E_{m}} \alpha_v  T_v^{(m)}
$$ 
and Eq.~(\ref{eq:Phi-map}) implies that
\begin{equation}\label{eq:Phi-special}
\Phi_{J,r}(X) = X\otimes \Pi_{J,r} = \frac{1}{2^n} \sum_{v+u\in E_{m}+ J} (-1)^{r(u)}\alpha_v   T_{v+u}^{(n)}.
\end{equation}  
The first step is to show that $\Phi_{J,r}$ maps $\Lambda_{m}$ into $\Lambda_n$. 
 Given a value assignment $s:I\to \ZZ_2$ and a subspace $J\subset E_n$ we write $s|_{I\cap J}$ for the restriction of $s$ to the intersection $I\cap J$. Given another value assignment $s':I\to \ZZ_2$ the delta function $\delta_{s',s}$ takes the value $1$ if $s'=s$ as functions on $I$ and $0$ otherwise.

\begin{Lemma}\label{lem1}
 Let $I$ be a maximal isotropic subspace of $E_n$. Then
$$ 
\Tr(\Phi_{J,r}(X) \Pi_{I,s}) =  \delta_{r|_{I\cap J},s|_{I\cap J}}  \frac{|I\cap J^\perp|}{2^n} \Tr(X\Pi_{I\cap E_{m},s|_{I \cap E_{m}} }) .
$$
In particular, if $X\in \Lambda_{m}$ then $\Phi_{J,r}(X)\in \Lambda_{n}$.
\end{Lemma}
{\em{Proof of Lemma~\ref{lem1}.}} We have $\Tr(\Phi_{J,r}(X))=1$ since $\Tr(X)=1$. We calculate  
$$
\begin{aligned}
\Tr(\Phi_{J,r}(X) \Pi_{I,s}) & =  \frac{1}{2^n} \sum_{v+u\in E_{m}+ J} (-1)^{r(u)}\alpha_v     \Tr( T_{v+u}    \Pi_{I,s}) \\
&  =  \frac{1}{2^n} \sum_{v+u\in E_{m}+ J} (-1)^{r(u)}\alpha_v   \frac{1}{|I|} \sum_{w\in I} (-1)^{s(w)}  \Tr( T_{v+u} T_w)  \\
& = \frac{|I\cap J|}{2^n} \sum_{v\in I\cap E_{m}} \alpha_v (-1)^{s(v)} \left( \frac{1}{|I\cap J|} \sum_{u\in I\cap J} (-1)^{r(u)}(-1)^{s(u)} \right)\\
&= \delta_{r|_{I\cap J},s|_{I\cap J}}  \frac{|I\cap J|}{2^n} \sum_{v\in I\cap E_{m}} \alpha_v (-1)^{s(v)}\\
&= \delta_{r|_{I\cap J},s|_{I\cap J}}  \frac{|I\cap J||I\cap E_{m}|}{2^n} \Tr(X\Pi_{I\cap E_{m},s|_{I \cap E_{m}}})\\
&= \delta_{r|_{I\cap J},s|_{I\cap J}}  \frac{|I\cap J^\perp|}{2^n} \Tr(X\Pi_{I\cap E_{m},s|_{I \cap E_{m}}})\geq 0.
\end{aligned}
$$ 
As a consequence of this calculation and the definition of $\Lambda_{n}$ the image $\Phi_{J,r}(X)$ belongs to $\Lambda_n$ for any $X\in \Lambda_{m}$.
$\Box$

Next we show that if $X\in \Lambda_{m}$ is a vertex then $\Phi_{J,r}(X)$ is a vertex of $\Lambda_n$. To achieve this we will use the following characterization of vertices of a polytope: 
\vspace{2mm}

{\it Let $P$ be a polytope in  $\RR^N$. Then a point $p\in P$ is a vertex if and only if there exists no non-zero $x\in \RR^N$ such that $p\pm x\in P$ \cite[page 18]{Grun13}.}
\vspace{2mm}

 Our polytope $\Lambda_n$ lives inside $\hH_1(2^n)$    and we can identify $\hH_1(2^n)$ with $\RR^{2^{2n}-1}$ by choosing $\id_{2^n}/2^n$ to be the origin. Therefore to prove that $\Phi_{J,r}(X) \in \Lambda_n$ is a vertex we need to show that 
 $\set{\Phi_{J,r}(X)\pm Y}\not\subset \Lambda_n$ for any  $Y\in \hH(2^n)$ of trace zero.
 Let us write
$$
Y=\frac{1}{2^n} \sum_{0\neq v\in E_{n}} \beta_v T_v^{(n)} 
$$
and define another trace zero operator 
\begin{equation}\label{eq:tildeY}
\tilde Y = \frac{1}{2^{m}} \sum_{0\neq v \in E_{m}} \tilde\beta_v T_v^{(m)}.
\end{equation}
The coefficients  $\tilde \beta_v$  are given by $\Tr(Y_v \Pi_{J,r})$ where $Y_v= 1/2^{n} \sum_{u\in J} \beta_{u+v} T_u^{(n)} \in \hH(2^n)$.  
Furthermore, we have the relation
\begin{equation} \label{rel1}
\tilde\beta_v = \frac{1}{|J|} \sum_{u\in J} \beta_{u+v} (-1)^{r(u)}.
\end{equation}
 
\begin{Lemma}\label{lem2}
Let $I'\subset E_{m}$ be a maximal isotropic subspace and $s':I'\to \ZZ_2$ be a value assignment. Then
\begin{equation}\label{eq:tildeY_tr}
\Tr(Y\Pi_{J+I',r\ast s'}) = \Tr(\tilde Y \Pi_{I',s'})
\end{equation}
 where $r\ast s': (J+I') \to \ZZ_2 $ is the value assignment defined by $r\ast s'(u+v) = r(u)+s'(v)$.
\end{Lemma}
{\em{Proof of Lemma~\ref{lem2}.}}
We calculate
$$
\begin{aligned}
\Tr(Y\Pi_{J+I',r\ast s'}) & = \frac{1}{|J||I'|} \sum_{v\in I'} \sum_{u\in J}  \beta_{u+v} (-1)^{r(u)+s(v)} \\
& = \frac{1}{|I'|} \sum_{v\in I'} (-1)^{s(v)}   \frac{1}{|J|} \sum_{u\in J}  \beta_{u+v} (-1)^{r(u)} \\
 & = \frac{1}{|I'|} \sum_{v\in I'} (-1)^{s(v)}   \tilde\beta_v \\
& =  \Tr(\tilde Y \Pi_{I',s'}).
\end{aligned}
$$
 For the third equality  we use Eq.~(\ref{rel1}). 
$\Box$

 We are now ready to give a proof of our main result, Theorem \ref{thm:Phi-map}.
\vspace{2mm}

\noindent
{\em{Proof of Theorem~\ref{thm:Phi-map}.}}
We begin by proving   Theorem \ref{thm:Phi-map} for $(J,r)=(J_0,r_0)$ in which case the map is given by Eq.~(\ref{eq:Phi-special}). Part (1)  follows immediately from   comparing  $\Pi_{J_0,r_0}Y\Pi_{J_0,r_0}$ after normalization with Eq.~(\ref{eq:Phi-special}).
For part (2) note that Lemma \ref{lem1} says that $\Phi_{J,r}(\Lambda_{m})\subset \Lambda_n$.
Let $X$ be a vertex of $\Lambda_{m}$. We will show that $\set{\Phi_{J,r}(X)\pm Y } \not\subset \Lambda_n$ for any   $Y\in \hH(2^n)$ of trace zero. Since $X$ is a vertex either $X+\tilde Y$ or $X-\tilde Y$ lies outside of $\Lambda_{m}$, where  $\tilde Y\in\hH(2^{m})$ is defined in Eq.~(\ref{eq:tildeY}). 
WLOG assume that $X+\tilde Y \not\in \Lambda_{m}$. This means that there is a pair $(I',s')$ such that $\Tr((X+\tilde Y)\Pi_{I',s'})<0$. 
  Lemma \ref{lem1} with $I=J+I'$ and $s=r\ast s'$ gives 
\begin{equation}\label{eq1}
\Tr(\Phi_{J,r}(X) \Pi_{J+I',r\ast s'}) = \Tr(X \Pi_{I',s'}).
\end{equation}
Using Eq.~(\ref{eq1}) and Eq.~(\ref{eq:tildeY})  we obtain
$$
\Tr((\Phi_{J,r}(X)+Y)\Pi_{J+I',r\ast s'}) = \Tr((X+\tilde Y)\Pi_{I',s'})<0,
$$
thus $X+Y$ lies outside of $\Lambda_n$. This completes the proof of part (2). Next we prove part (3). If $X=A_\Omega^\gamma$ for some cnc set $(\Omega,\gamma)$ where  $\Omega \subset E_{m}$ then Eq.~(\ref{eq:Phi-special}) implies that 
$$ 
\Phi_{J,r}(A_\Omega^\gamma)  =  \frac{1}{2^n} \sum_{u+v\in J+ E_{m}} (-1)^{r(u)+\gamma(v)}  T_{u+v}  = A_{J+\Omega}^{r\ast \gamma}.
$$
Therefore we obtain a vertex of cnc-type. Conversely, if $\Phi_{J,r}(X)$ is a vertex of cnc-type given by $A_{\tilde\Omega}^{\tilde\gamma }$
for some cnc set $(\tilde\Omega,\tilde\gamma)$ then again from Eq.~(\ref{eq:Phi-special}) we see that $X=A_\Omega^\gamma$ where $\Omega = \tilde \Omega \cap E_{m}$ and $\gamma$   is given by the restriction $\tilde \gamma|_{\Omega}$. 

For the general case let $J\subset E_n$ be an arbitrary isotropic subspace and $r:J\to \ZZ_2$ be a value assignment. We will write $J_0$ for $\Span{x_{m+1},\cdots,x_n}$ and $r_0:J_0\to \ZZ_2$ for the value assignment defined by $r_0(v)=0$ for all $v\in J_0$.
Let $U$ be a Clifford unitary as described in the statement of the Theorem. We think of $\Phi_{J,r}$ as the composite of two maps:
$$ \Phi_{J,r}(X) =U\Phi_{J_0,r_0}(X) U^\dagger.$$ 
Part (1) follows from the relation $U \Pi_{J_0,r_0}U^\dagger = \Pi_{J,r}$. 
Part (2) holds since $U$ acts on $\Lambda_n$ by permuting its vertices \cite{HVM}. Also this action maps a cnc-type vertex to a cnc-type vertex, which implies part (3). $\Box$

\section{Reduction of the classical simulation}\label{sim}

We now turn to the second question posed in the introduction, namely whether the classical simulation of the update of the composite vertices---whose existence was established in Theorem~\ref{thm:Phi-map}---under the operations of QCM can be reduced to the classical simulation of their constituent parts. This is indeed the case, as Theorem~\ref{T1} below demonstrates.

\begin{Theorem}\label{T1}
Any quantum computation in the magic state model (QCM) that operates on an initial state $U(X_A \otimes (\Pi_\sigma)_B)U^\dagger$, where $X\in \Lambda_m$, is an $m$-qubit  vertex and $\Pi_\sigma:=|\sigma\rangle\langle \sigma|$ is the projector on an $(n-m)$-qubit stabilizer state $|\sigma\rangle$ and $U$ is an $n$-qubit Clifford unitary, can be efficiently reduced to a QCM on initial state $X$ alone.
\end{Theorem}
The theorem says that supplementing a vertex $X$ with a stabilizer state does not increase the computational power of QCM.   A proof of a similar result is given in Ref.~\cite[\S V]{BravyiSmolin2016} for the case where $X_A$ in the statement of Theorem~\ref{T1} is replaced by the state $|T\rangle\langle T|^{\otimes m}$.\medskip

{\em{Proof of Theorem~\ref{T1}.}} We start from the version of QCM where the quantum computation consists of a sequence of Pauli measurements. All Clifford unitaries can be propagated forward past the last measurement (conjugating the measured observables in the passing), and then discarded. Thus, wlog. we consider initial states of form $\tilde{X} =X_A \otimes (\Pi_\sigma)_B$.

We give an explicit procedure to replace the sequence ${\cal{T}}$ on $A \otimes B$ by an equivalent sequence $\tilde{T}^{(A)}$ of observables that act only on the subsystem $A$. The proof is by induction, and the induction hypothesis is that, at time $t$, the sequence ${\cal{T}}_{\leq t}$ of measurements has been replaced by a computationally equivalent sequence $\tilde{\cal{T}}^{(A)}_{\leq t}$ of Pauli measurements on the register $A$ only. This statement is true for $t=0$, i.e., the empty measurement sequence. We now show that the above statement for time $t$ implies the analogous statement for time $t+1$. 

At time $t$, the state of the quantum register evolved under the computationally equivalent measurement sequence $\tilde{\cal{T}}^{(A)}_{\leq t}$ is $\tilde{Y}(t) = Y(t)_A\otimes (\Pi_\sigma)_B$. We now consider the Pauli observable $T(t+1)\in {\cal{T}}$ to be measured next, and write $T(t+1)= R_A(t+1) \otimes S_B(t+1)$. There are two cases:\smallskip 

Case I: {\em{$T(t+1)$ commutes with the entire stabilizer ${\cal{S}}$ of $|\sigma\rangle$.}} Hence, also $S_B(t+1)$ commutes with ${\cal{S}}$. But then, either $S_B(t+1)$ or $-S_B(t+1)$ is in ${\cal{S}}$, and $S_B(t+1)$ may be replaced by its eigenvalue $\pm 1$ in the measurement. Hence, the measurement of $T(t+1)$ is equivalent to the measurement of $\pm R_A(t+1)$.\smallskip

Case II: {\em{$T(t+1)$ does not commute with the entire stabilizer ${\cal{S}}$ of $|\sigma\rangle$.}} Then, the measurement outcome $s_{t+1}$ is completely random. Further, there exists a Clifford unitary $V$ such that
$$ 
V {\cal{S}}(t+1) V^\dagger = \langle X_{B:1}, X_{B:2},..,X_{B:m}\rangle,\;\; V T(t+1) V^\dagger = Z_{B:1}.
$$
Therefore, the state resulting from the measurement of $T(t+1)$, with outcome $s_{t+1}$ on the state $\tilde{Y}(t)$ is the same state as the one resulting from the following procedure: 
\begin{enumerate}
\item{Apply the Clifford unitary $V$ to $\tilde{Y}(t)  = Y(t)_A\otimes (\Pi_\sigma)_B$, leading to
$$
V\tilde{Y}(t) V^\dagger  = Y'(t) \otimes |\overline{+}\rangle\langle \overline{+} |_B,
$$
where $|\overline{+}\rangle_B:=\bigotimes_{i\in B}|+\rangle_{B:i}$.}
\item{Measure $Z_{B:1}$ on $Y'(t) \otimes |\overline{+}\rangle\langle \overline{+} |_B$, with outcome $s_{t+1}$.}
\item{Apply $V^\dagger$.}
\end{enumerate}
Now, note that the measurement in Step 2, of the Pauli observable $Z_{B:1}$ is applied to the stabilizer state $|\overline{+}\rangle_B$. The result is $ \sigma'(t+1)\rangle = |s_{t+1}\rangle_{B:1} \bigotimes_{j=2}^m |+\rangle_{B:j}$; that is the first qubit of subsystem $B$ is now in a $Z$-eigenstate, and the other qubits are unchanged. Therefore, after normalization, the effect of the measurement can be replaced by the unitary $\left(X_{B:1}\right)^{s_{t+1}} H_{B:1}$.

Thus, the whole procedure may be replaced by the Clifford unitary $V^\dagger \, \left(X_{B:1}\right)^{s_{t+1}} H_{B:1} \, V$. But Clifford unitaries don't need to be implemented. They are just propagated past the last measurement, thereby affecting the measured observables by conjugation whereby their Pauli-ness is preserved. In result, in Case II, the measurement of $T(t+1)$ doesn't need to be performed at all. It is replaced by a coin flip, and efficient classical post-processing of the subsequent measurement sequence.\medskip

We conclude that in both the cases I and II, given the induction assumption, the original measurement sequence ${\cal{T}}_{\leq t+1}$ can be replaced by a computationally equivalent measurement sequence $\tilde{\cal{T}}^{(A)}_{\leq t+1}$ acting on register $A$ only. By induction, the complete measurement sequence ${\cal{T}}$ can be replaced by a computationally equivalent sequence $\tilde{\cal{T}}^{(A)}$ acting on $A$ only. 

Since the measurements $\tilde{\cal{T}}^{(A)}$ are applied to an unentangled initial state $X_A \otimes (\Pi_\sigma)_B$, the register $B$ may be dropped without loss of information. Thus, any sequence of Pauli measurements on the initial state $X_A\otimes (\Pi_\sigma)_B$ is efficiently reduced to a Pauli measurement sequence of at most the same length on $X$ alone. $\Box$

\paragraph{Discussion.} Theorem~\ref{T1} extends the previously known range of efficient classical simulation of QCM by sampling from the polytopes $\Lambda_n$. The prior result is Theorem~3 from \cite{raussendorf2020phase}. It says that   if a quantum state $\rho$ can be represented as a probabilistic linear combination of cnc vertices, and the probability distribution defining this expansion can be efficiently sampled from, then any QCM on $\rho$ can be efficiently classically simulated. Classical simulation of probability distributions over stabilizer states is contained therein as a limiting case. 

To apply the above Theorem~\ref{T1}, we define vertex classes  
$$\wW_K:=\{U(A_\alpha \otimes \Pi_{\sigma})U^\dagger |\, A_\alpha \in \Lambda_K, |\sigma\rangle\in\mathcal{S}_{n-K}, U\in\Cl_n, n\geq K\}$$
consisting of a union of certain $n$-qubit vertices for $n\geq K$.
 Therein, $\Pi_{\sigma}$ is the projector corresponding to a $(n-K)$-qubit stabilizer state $|\sigma\rangle$ and $\Cl_n$ is the $n$-qubit Clifford group.
Every class  $\wW_K$ contains of vertices on arbitrarily many qubits  for $n\geq K$. The present extension of Theorem~3 of \cite{raussendorf2020phase}, based on Theorem~\ref{T1} above, is the following.
\begin{Cor}\label{Co1}
Consider an $n$-qubit quantum state that can be expanded into a probabilistic linear combination of vertices from the set $\wW_K$, for a given value of $K$. Then, the computational cost of simulating any QCM on $\rho$ is polynomial in $n$.
\end{Cor}
Note that the simulation is (likely to be) inefficient in $K$. 

\section{Beyond vertices of cnc-type}\label{BeyondCNC}  

 In the previous section we established that the complexity of simulating QCM for a composite vertex $U(X_A \otimes (\Pi_\sigma)_B)U^\dagger$ efficiently reduces to the simulation of QCM on $X_A$. But which vertices $X$ can we actually put explicitly into this reduction?---To date, the only family of vertices which is explicitly described, including the update under Pauli measurements, are the vertices of cnc-type. 

For a single qubit, all eight vertices are equivalent under Clifford transformations, and cnc. For two qubits, there are $8$ Clifford-equivalence classes of vertices, two of which are of cnc-type. 

In this section, we give a complete characterization of one class of two-qubit vertices that are not cnc.
A distinguishing feature of the new type of vertices is that the expectations $\Span{T_v}$ take values in $\set{0,\pm 1/2,\pm 1}$ where as in the cnc case these expectations belong to $\set{0,\pm 1}$.

Our goal is to describe the Clifford orbit of the vertex denoted by $A_{\alpha_0}$ whose coordinates in the Pauli basis (i.e. the expectations $\Span{T_v}$)  are given as follows 
$$
\begin{array}{cccccccccccccccc}
II & IX & XI & XX & IZ & IY & XZ & XY & ZI & ZX & YI & YX & ZZ & ZY & YZ & YY  \\
 1 &  -1/2 &   1/2 &     0&  -1/2&  -1/2 &    -1&     0&   1/2&    -1&  -1/2&     0&     0&     0&    0&     1 
\end{array}
$$ 
  We know that there are $1920$ vertices in this orbit (by computer calculation). The set of vertices in the Clifford orbit of $A_{\alpha_0}$ will be denoted by
\begin{equation}\label{eq:ClOrbit}
\oO = \set{UA_{\alpha_0}U^\dagger|\; U\in\Cl_n }.
\end{equation}
For our construction  we will consider  noncontextual subsets of $E_2$ that are not necessarily closed.
Let $\Omega$ be a subset of $E_n$ and $\lambda:\Omega\to \ZZ_2$ be a value assignment.  Associated to $(\Omega,\gamma)$ we can define the operator   $A_\Omega^\gamma$ as in Eq.~(\ref{eq:cnc-type}).  

\begin{figure}[!htb]
    \centering
    \resizebox{0.7\textwidth}{!}{
     
\centering
\begin{tikzpicture} 
\node[circle, fill=blue!30] (XI) {$XI$};
\node[circle, fill=red!50] (XI-IZ) [above  right of = XI, right=2cm, above=0.125cm] {};
\node[circle, fill=red!50] (XI-IX) [above left of=XI,left=2cm, above=0.125cm] {};
\node[circle, fill=red!50] (XI-IY) [below of=XI, below =0.5cm] {};
\node[circle, fill=blue!30] (XZ) [above of =XI-IZ,above=0.5cm] {$XZ$};
\node[circle, fill=purple!100] (XZ-ZY) [above right of = XZ] {};
\node[circle, fill=purple!100] (XZ-ZX) [above left of = XZ] {};
\node[circle, fill=blue!30] (ZY1) [above  of = XZ-ZY] {$ZY$};
\node[circle, fill=blue!30] (YX1) [above right of = XZ-ZY] {$YX$};
\node[circle, fill=blue!30] (YY1) [above of = XZ-ZX] {$YY$};
\node[circle, fill=blue!30] (ZX1) [above left of = XZ-ZX] {$ZX$};
\node[circle, fill=blue!30] (IX) [above of =XI-IX,above=0.5cm] {$IX$}; 
\node[circle, fill=red!50] (IX-ZI) [above right of = IX] {};
\node[circle, fill=blue!10] (ZI3) [above of = IX-ZI] {$ZI$}; 
\node[circle, fill=blue!10] (ZX3) [above right of = IX-ZI] {$ZX$}; 
\node[circle, fill=red!50] (IX-YI) [above left of = IX] {}; 
\node[circle, fill=blue!10] (YX3) [above  of = IX-YI] {$YX$}; 
\node[circle, fill=blue!10] (YI3) [above left of = IX-YI] {$YI$};
\node[circle, fill=blue!30] (XX) [left of =XI-IX,left=0.5cm] {$XX$};  
\node[circle, fill=purple!100] (XX-YZ) [above left of = XX] {};
\node[circle, fill=blue!10] (ZY3) [above of = XX-YZ] {$ZY$}; 
\node[circle, fill=blue!10] (YZ3) [above left of = XX-YZ] {$YZ$}; 
\node[circle, fill=purple!100] (XX-YY) [below left of = XX] {}; 
\node[circle, fill=blue!10] (ZZ3) [left of = XX-YY] {$ZZ$}; 
\node[circle, fill=blue!10] (YY3) [below left of = XX-YY] {$YY$};   
\node[circle, fill=blue!30] (IZ) [right of =XI-IZ,right=0.5cm] {$IZ$};
\node[circle, fill=red!50] (IZ-ZI) [below right of = IZ] {};
\node[circle, fill=red!50] (IZ-YI) [above right of = IZ] {}; 
\node[circle, fill=blue!30] (YZ1) [above right of = IZ-YI] {$YZ$};
\node[circle, fill=blue!30] (YI1) [right of = IZ-YI] {$YI$};
\node[circle, fill=blue!30] (ZZ1) [right of = IZ-ZI] {$ZZ$};
\node[circle, fill=blue!30] (ZI1) [below right of = IZ-ZI] {$ZI$};
\node[circle, fill=blue!30] (IY) [below left of = XI-IY,below=0.25cm,left=0.75cm] {$IY$};
\node[circle, fill=red!50] (IY-YI) [left of =IY] {};
\node[circle, fill=blue!20] (YI2) [above left of = IY-YI] {$YI$};
\node[circle, fill=blue!20] (YY2) [below left of = IY-YI] {$YY$};
\node[circle, fill=red!50] (IY-ZI) [below of = IY] {};
\node[circle, fill=blue!20] (ZI2) [below left of = IY-ZI] {$ZI$};
\node[circle, fill=blue!20] (ZY2) [below right of = IY-ZI] {$ZY$};
\node[circle, fill=blue!30] (XY) [below right of = XI-IY,below=0.25cm,right=0.75cm] {$XY$};
\node[circle, fill=purple!100] (XY-YZ) [ right of =XY] {};
\node[circle, fill=blue!20] (YZ2) [above right of = XY-YZ] {$YZ$};
\node[circle, fill=blue!20] (ZX2) [below right of = XY-YZ] {$ZX$};
\node[circle, fill=purple!100] (XY-YX) [below of = XY] {};
\node[circle, fill=blue!20] (YX2) [below left of = XY-YX] {$YX$};
\node[circle, fill=blue!20] (ZZ2) [below right of = XY-YX] {$ZZ$};
 
\path (XI) edge  (XI-IX);
\path (XI-IX) edge  (IX); 
\path (IX) edge (IX-ZI); 
\path (IX-ZI) edge (ZX3); 
\path (IX) edge (IX-YI);
\path (XI-IX) edge  (XX);
\path (XX) edge (XX-YZ);
\path (XX) edge (XX-YY);
\path (XI) edge  (XI-IZ); 
\path (XI-IZ) edge  (XZ); 
\path (XZ) edge (XZ-ZX);
\path (XZ-ZX) edge (ZX1);
\path (XZ) edge (XZ-ZY); 
\path (XI-IZ) edge  (IZ);
\path (IZ) edge (IZ-YI);
\path (IZ) edge (IZ-ZI);
\path (XI) edge  (XI-IY);
\path (XI-IY) edge  (IY); 
\path (IY) edge (IY-YI);
\path (IY) edge (IY-ZI);
\path (XI-IY) edge  (XY);
\path (XY) edge (XY-YX);
\path (XY) edge (XY-YZ);
\path (XY-YZ) edge (ZX2);
\path (XZ-ZX) edge (YY1);
\path (IY-YI) edge (YY2);
\path (XX-YY) edge (YY3);
\path (XZ-ZY) edge (ZY1); 
\path (IY-ZI) edge (ZY2); 
\path (XX-YZ) edge (ZY3);  
\path (XZ-ZY) edge (YX1);
\path (XY-YX) edge (YX2);
\path (IX-YI) edge (YX3);
\path (IZ-YI) edge (YZ1);
\path (XY-YZ) edge (YZ2);
\path (XX-YZ) edge (YZ3);
\path (IZ-YI) edge (YI1);
\path (IY-YI) edge (YI2);
\path (IX-YI) edge (YI3);
\path (IZ-ZI) edge (ZZ1);
\path (XY-YX) edge (ZZ2);
\path (XX-YY) edge (ZZ3);
\path (IZ-ZI) edge (ZI1);
\path (IY-ZI) edge (ZI2);
\path (IX-ZI) edge (ZI3); 
\end{tikzpicture}
    }
    \caption{The poset of  isotropic subspaces of $E_2$. Large blue nodes correspond $1$-dimensional subspaces and small red nodes correspond to $2$-dimensional subspaces.   Each node at the boundary repeats $3$ times and they are identified.  
 The dark red colored nodes represent the maximal isotropic subspaces in the collection $\cC$ corresponding to the vertex $A_{\alpha_0}$.   
} \label{fig:isoptropics}
\end{figure}
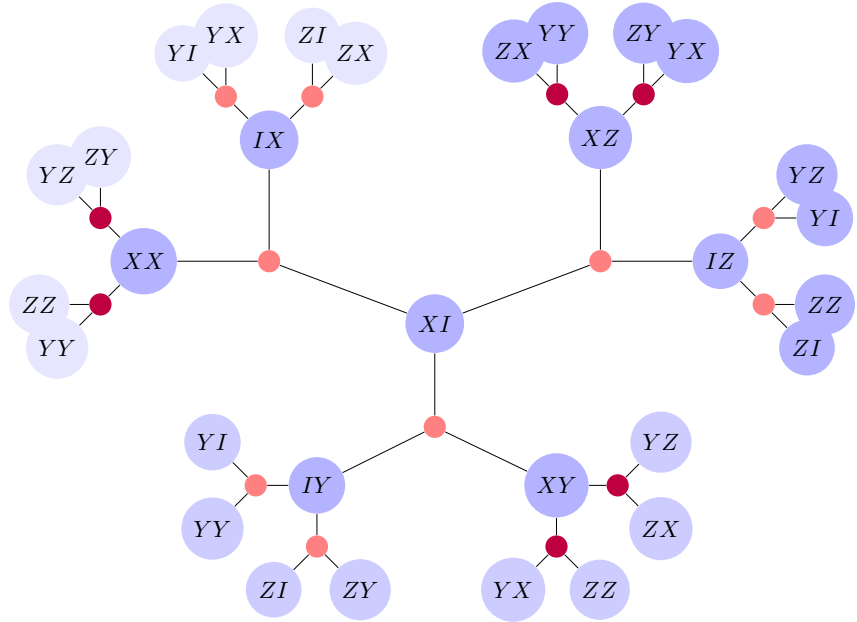

\paragraph{Construction.}  The vertices in the Clifford orbit $\oO$ given in Eq.~(\ref{eq:ClOrbit}) have the following form 
\begin{equation}\label{eq:AIOmega}
A_{I,\Omega}^\gamma = A_I^\gamma + \frac{1}{4}(A_\Omega^{\gamma'}- A_\Omega^{\gamma''} ).
\end{equation} 
where 
\begin{enumerate}
\item[C.1] $I\subset E_2$ is a maximal isotropic subspace.

\item[C.2] $\gamma:I\to \ZZ_2$ is a value assignment.

\item[C.3] The set $\Omega$ is constructed from a collection $\cC$ described in (C.4) by the following formula
\begin{equation}\label{eq:Omega-C}
 \Omega =  E_2 -  \Omega^\perp \;\;\;\text{ where }\;\; \Omega^\perp=  \left( \bigcup_{J\in \cC} J \right) - \set{0}.
\end{equation}

\item[C.4] The collection $\cC=\cC(I)$ of maximal isotropics (see Fig.~(\ref{fig:isoptropics})) is defined using the following rules
\begin{enumerate}
\item[R.1] $\cC$ contains $I$,
\item[R.2]   for each $J\in \cC$ and  $0\neq v\in J $ exactly one of the two subspaces $J'\in \iI(E_2)-\set{J}$ containing $v$ is contained in $\cC$. 
\end{enumerate}

\item[C.5] The value assignments $\gamma'$ and $\gamma''$ on $\Omega$ are uniquely specified by the requirements   
\begin{enumerate}
\item[V.1] $\gamma'(0)=\gamma''(0)=0$,
\item[V.2] $\gamma'(v)=1+\gamma''(v)$ for $v\in \Omega-\set{0}$,
\item[V.3] $\gamma'(v)+\gamma'(w)+\beta(v,w)=\gamma(v+w)$ for $v,w\in \Omega$ such that $[v,w]=0$ and $v+w\in I$. Note that (V.2) implies that $\gamma''$ also satisfies this property.
\end{enumerate}  
\end{enumerate}
In summary a vertex in the Clifford orbit $\oO$ is specified by $(I,\gamma,\cC)$. Counting these components: $15$ maximal isotropics, $2^2$ value assignments, $2^5$ collections $\cC$  we obtain a total of $1920=15 \times  2^7$ vertices. This covers all the vertices in the Clifford orbit of $A_{\alpha_0}$.

The vertex $A_{\alpha_0}$ has the following parameters: $I=\Span{x_1+z_2,z_1+x_2}$, $\cC$ consists of the following collection
$$\set{\Span{x_1+z_2,z_1+x_2},\Span{x_1+x_2,z_1+z_2},\Span{x_1+y_2,y_1+z_2},\Span{z_1+y_2,y_1+x_2},\Span{y_1+x_2,x_1+y_2},\Span{z_1+y_2,y_1+z_2} },$$ 
and  the value assignments are given by
$$
\begin{array}{ccccccccccccccccc}
&II & IX & XI & XX & IZ & IY & XZ & XY & ZI & ZX & YI & YX & ZZ & ZY & YZ & YY  \\
(-1)^\gamma & 1 &   &    &     &  &   &    -1&     &   &    -1&  &     &   &     &    &     1 \\
(-1)^{\gamma'}  & 1 &  -1 &  1  &     &  -1  & -1  &    &   &1   &    & -1 &     &   &     &    &   \\
(-1)^{\gamma''}  & 1 &  1 &  -1  &     &  1  & 1  &    &   &-1   &    & 1 &     &   &     &    &       
\end{array}
$$ 
The set $\Omega$ can be calculated from $\cC$ using Eq.~(\ref{eq:Omega-C}) and it turns out to be $\set{0,x_1,y_1,z_1,x_2,y_2,z_2}$. 

 Our strategy for computing the update rules for $A_{I,\Omega}^\gamma$ is to consider the update of $A_I^\gamma$ and $\frac{1}{4}(A_\Omega^{\gamma'}-A_\Omega^{\gamma''})$  in Eq.~(\ref{eq:AIOmega}) separately. First one can be updated using the update rule  \cite{raussendorf2020phase} of cnc-type vertices:
\begin{equation}\label{eq:CNC-update}
\arraycolsep=1.4pt\def\arraystretch{2.2}
\Pi_{a,s} A_{I}^\gamma \Pi_{a,s} =\left\lbrace
\begin{array}{cc}
\delta_{s_a,\gamma(a)} A_I^\gamma & a\in I \\
\frac{1}{2} A_{I\times a}^{\gamma\times a} & a\notin I.   
\end{array}
\right.
\end{equation}
For the second one we   first define a cnc set $\tilde \Omega$ obtained as the closure of $\Omega \cap \Span{a}^\perp$ under inference. We also extend $\gamma'$ and $\gamma''$ on this cnc set in the expected way. The resulting value assignments are denoted by $\tilde\gamma'$ and $\tilde\gamma''$, respectively. Then the update of $\frac{1}{4}(A_\Omega^{\gamma'}-A_\Omega^{\gamma''})$ coincides with the update of $\frac{1}{4}(A_{\tilde\Omega}^{\tilde\gamma'}-A_{\tilde\Omega}^{\tilde\gamma''})$ by the properties of $\gamma'$ and $\gamma''$. By construction $\tilde \Omega$ is contained in $\Span{a}^\perp$, in particular it contains $a$.
Eq.~(\ref{eq:CNC-update}) gives us
$$\arraycolsep=1.4pt\def\arraystretch{2.2}
\Pi_{a,s}  A_{\tilde\Omega}^{\tilde\gamma'} \Pi_{a,s} = \delta_{s_a,\tilde\gamma'(a)} A_{\tilde \Omega}^{\tilde\gamma'}.
$$
Replacing $\tilde\gamma'$ with $\tilde\gamma''$ gives the update rule for $A_{\tilde \Omega}^{\tilde \gamma''}$. 
 Next we give the update rules for $A_{I,\Omega}^\gamma$. The proof of this result is given in Appendix \ref{sec:Proof-Update}.

\begin{Theorem}\label{thm:update-rules}
 The update rules for $A_{I,\Omega}^\gamma$ under Pauli measurements consist of the following three cases:
\begin{enumerate}[leftmargin=3\parindent]

\item[Case I.] Suppose that   $a\in I$ and $a\notin \Omega$.  \\
\begin{enumerate}
\item[(1.a)] If  $\gamma(a)\neq s$ then $\Pi_{a,s} A_{I,\Omega}^\gamma \Pi_{a,s}=0.$ \\
\item[(1.b)] If $\gamma(a)=s$   then
$$
\frac{\Pi_{a,s} A_{I,\Omega}^\gamma \Pi_{a,s} }{\Tr(A_{I,\Omega}^\gamma \Pi_{a,s})} =     \frac{2A_{\Span{a}^\perp}^{\tilde\alpha_0} +A_{\Span{a}^\perp}^{\tilde\alpha_1} +A_{\Span{a}^\perp}^{\tilde\alpha_2} }{4} 
$$   
where the value assignments are given in Eq.~(\ref{eq:1b}).
\end{enumerate}

\item[Case II.] Suppose that   $a\notin I$ and $a\in \Omega$.\\
\begin{enumerate}
\item[(2.a)] If $\tilde\gamma'(a)\neq s$ then
$$ 
\frac{\Pi_{a,s} A_{I,\Omega}^\gamma \Pi_{a,s} }{\Tr(A_{I,\Omega}^\gamma \Pi_{a,s})}   =  A_{\Span{a}^\perp}^{\tilde \alpha'}
$$ 
where the value assignments are given in Eq.~(\ref{eq:2a}).
\item[(2.b)] If $\tilde\gamma'(a)= s$ then
$$
\frac{\Pi_{a,s} A_{I,\Omega}^\gamma \Pi_{a,s} }{\Tr(A_{I,\Omega}^\gamma \Pi_{a,s})}  =   \frac{2A_{\Span{a}^\perp}^{\tilde \alpha} + A_{\Span{a}^\perp}^{\tilde\alpha'} }{3} 
$$ 
where the value assignments are given in Eq.~(\ref{eq:2b}).
\end{enumerate}

\item[Case III.] Suppose that  $a\notin I$ and $a\notin \Omega$.\\
\begin{enumerate}
\item[(3.a)] If $\tilde\gamma'(a)\neq s$ then
$$
\frac{\Pi_{a,s} A_{I,\Omega}^\gamma \Pi_{a,s} }{\Tr(A_{I,\Omega}^\gamma \Pi_{a,s})}  =   \frac{A_{\Span{a}^\perp}^{\tilde \alpha_0}+ A_{\Span{a}^\perp}^{\tilde \alpha_1}  }{2}
$$ 
where the value assignments are given in Eq.~(\ref{eq:3a}).
\item[(3.b)] If $\tilde\gamma'(a)= s$ then
$$
\frac{\Pi_{a,s} A_{I,\Omega}^\gamma \Pi_{a,s} }{\Tr(A_{I,\Omega}^\gamma \Pi_{a,s})}   =     \frac{A_{\Span{a}^\perp}^{\tilde\alpha_0} +  A_{\Span{a}^\perp}^{\tilde\alpha_1}}{2} 
$$   
where the value assignments are given in Eq.~(\ref{eq:3b}).
\end{enumerate}
\end{enumerate}
\end{Theorem}


\paragraph{Discussion.}  Prior to this work the only class of $n$-qubit vertices that is known to 
 give efficient classical simulation was  the class of cnc vertices \cite{raussendorf2020phase}. 
  For the efficiency of the simulation each update rule is required to produce a probability distribution which can be  efficiently sampled.
Our description of the new type of vertices in $\Lambda_2$ and their update rules under Pauli measurements allows us to extend this  result to a larger class of vertices. Moreover, this is not restricted to $n=2$ case. Theorem \ref{T1} implies that the classically efficiently simulable sector of $\Lambda_n$ extends with these new classes of vertices. To make this precise (relying on Theorem \ref{thm:Phi-map}) we enlarge the set of cnc-type vertices by taking the disjoint union of $n$-qubit cnc vertices with the image  of the vertices in $\oO$ defined in Eq.~\ref{eq:ClOrbit} under the $\Phi$-map:
$$
\vV_n' := \vV_{n}^\cnc \sqcup  \set{U(A_\alpha \otimes \Pi_{\sigma})U^\dagger  |\;A_\alpha\in \oO}
$$
where $\Pi_{\sigma}$ is a $(n-2)$-qubit stabilizer projector and $U$ is an $n$-qubit  Clifford unitary. 

\begin{Cor}\label{Co2}
Consider an $n$-qubit quantum state $\rho$ that can be expressed as a probabilistic linear combination $p_\rho:\vV_n' \to \RR_{\geq 0} $. If $p_\rho$  can be efficiently sampled from then the classical simulation  of any QCM on $\rho$ is efficient. 
\end{Cor}

\section{Conclusion}\label{Concl} 

This paper is an expedition into the state polytopes $\Lambda_n$ \cite{HVM,Heim}, which are presently largely uncharted territory. The study of these polytopes is motived by the fact that they form the structural basis of a hidden-variable description of universal quantum computation \cite{HVM}. 

Here we have shown that certain extremal points of the $\Lambda_n$ can be built from ``smaller parts'', namely a vertex of state polytope $\Lambda_m$ on a smaller number $m<n$ of qubits, and a  stabilizer state. Further, we have shown that the classical simulation of the evolution of such composite vertices can be reduced to the evolution of their parts.   We have also described a new class of vertices outside the known cnc classification, together with their update rules under Pauli measurements.

A possible next question is as follows. The map $\Phi$ takes vertices of  $\Lambda_m$ to vertices of $\Lambda_n$, $n>m$, by tensoring on projectors onto stabilizer states. Are there generalizations of this map that use more general constructs than stabilizer states as parameters of the mapping, and more general notions of composition than the tensor product?
\smallskip

To conclude with a broader comment, the state polytopes $\Lambda_n$ are a novel object in the theory of quantum computation and foundations of quantum mechanics. It is presently unknown how easy or hard their study is going to be, and which techniques will be useful.  With the present investigation, we have made a first dent into the subject.

\paragraph{Acknowledgments.} CO is supported by the Air Force Office of Scientific Research under award number FA9550-21-1-0002.  RR acknowledges funding from NSERC, in part through the Canada First Research Excellence Fund, Quantum Materials and Future Technologies Program.

\appendix

\section{Proof of Theorem \ref{thm:update-rules} }\label{sec:Proof-Update}

\noindent As a preparation for the update rules of $A_{I,\Omega}^\gamma$ we consider certain linear combinations of operators in $\hH(2)$ and   express them  as probabilistic mixtures of the vertices of $\Lambda_1$. 
We will denote the vertices in $\Lambda_1$ by $A_E^\alpha$ where $\alpha$ is a value assignment on $E=E_1$. Other points in $\Lambda_1$ that we are interested in are the projectors $\Pi_{v,s_v}$ 
where $v=x,y,z$, where $y=x+z$, and $s_v=0,1$.
\medskip

\begin{Lemma}\label{lem:decomposition}We have the following decompositions for the points in $\Lambda_1$. Let $v,w\in E$ be non-zero distinct elements.
\begin{enumerate}
\item Let $\alpha_0$ and $\alpha_1$ be value assignments on $E$ defined by  $\alpha_0(v)=\alpha_1(v)=s_v$, $\alpha_1(w)=1+\alpha_0(w)$ and $\alpha_1(v+w)=1+\alpha_0(v+w)$. Then
$$
  \Pi_{v,s_v} = \frac{A_{E}^{\alpha_0}+A_{E}^{\alpha_1}}{2}.
$$

\item Let $\alpha_0$ and $\alpha_1$ be value assignments on $E$ defined by  $\alpha_0(v)=\alpha_1(v)=s_v$, $\alpha_0(w)=\alpha_1(w)=s_w$ and $\alpha_1(v+w)=1+\alpha_0(v+w)$. Then  
$$
\Pi_{v,s_v} + \frac{\Pi_{w,s_w}-\Pi_{w,s_w+1}}{2 }=\frac{A_{E}^{\alpha_0}+A_{E}^{\alpha_1}}{2}.
$$

\item  Let $\alpha'$  be value assignment on $E$ defined by  $\alpha'(v)=\alpha(v)$, $\alpha'(w)=1+\alpha(w)$ and $\alpha'(v+w)=1+\alpha(v+w)$. Then
$$
\frac{2 \Pi_{v,\alpha(v)} + A_E^\alpha}{3} = \frac{2A_E^\alpha + A_E^{\alpha'}}{3}.
$$

\item  Let $\alpha'$  be value assignment defined in (3). Then
$$
2\Pi_{v,\alpha(v)} - A_E^\alpha = A_E^{\alpha'}.
$$

\item Let $\alpha_i$, $i=0,1,2$, be defined by $\alpha_i(v)=s_v$ for all $i$, $\alpha_0(w)=\alpha_1(w)= 1+ \alpha_2(w)=s_w$, $\alpha_1(v+w)=\alpha_2(v+w)=1+\alpha_0(v+w)$. Then 
$$
\Pi_{v,s_v} + \frac{\Pi_{w,s_w}-\Pi_{w,s_w+1}}{ 4}= \frac{2A_{E}^{\alpha_0}+A_{E}^{\alpha_1}+A_{E}^{\alpha_2} }{4}.
$$
\end{enumerate}
\end{Lemma}
{\em{Proof of Lemma~\ref{lem:decomposition}.}}
Verification is straightforward: We write   each side of the equations in the Pauli basis and compare. $\Box$
 

Let us label the maximal isotropics in $\Span{a}^\perp$ by $I_v=\Span{\tilde v,a}$, $I_w=\Span{\tilde w,a}$ and $I_{v+w}=\Span{\tilde v + \tilde w,a}$.
We introduce a map
\begin{equation}\label{eq:vertex-map}
\Lambda_1 \to \Lambda_2
\end{equation}
determined by $A_{E}^\alpha \mapsto   A_{\Span{a}^\perp}^{\tilde \alpha}$ where $\tilde \alpha$ is specified by $\tilde \alpha(a)=s_a$, $\tilde \alpha(b)=\alpha(b)$ for $b=\tilde v,\tilde w, \tilde v+\tilde w$. 
In the following proof   we label the maximal isotropic subspaces of $\Span{a}^\perp$ by $I_v,I_w,I_{v+w}$.


\noindent {\em{Proof of Theorem~\ref{thm:update-rules}.}}
In Case I we have  
$$
\Pi_{a,s} A_{I,\Omega}^\gamma \Pi_{a,s}  = \delta_{s_a,\gamma(a)} \left( A_I^\gamma +\frac{1}{4} (A_{\tilde \Omega}^{\tilde \gamma'} -  A_{\tilde \Omega}^{\tilde\gamma''})  \right)
$$
where we used $\tilde\gamma'(a)=\tilde\gamma''(a)=\gamma(a)$ since $a=v+w$ for some $v,w\in\Omega-\set{0}$. In this case $I$ and $\tilde \Omega$ are maximal isotropics in $\Span{a}^\perp$ intersecting at $\Span{a}$. (1.a) follows immediately by calculating $\Pi_{a,s} A_{I,\Omega}^\gamma \Pi_{a,s}$. For (1.b) we observe that  Lemma \ref{lem:decomposition} part (5) and Eq.~(\ref{eq:vertex-map}) imply that
$$
\frac{\Pi_{a,s} A_{I,\Omega}^\gamma \Pi_{a,s} }{\Tr(A_{I,\Omega}^\gamma \Pi_{a,s})} =  A_I^\gamma +\frac{1}{4}(A_{\tilde \Omega}^{\tilde\gamma'}-A_{\tilde \Omega}^{\tilde \gamma''})  =   \frac{2A_{\Span{a}^\perp}^{\tilde\alpha_0} +A_{\Span{a}^\perp}^{\tilde\alpha_1} +A_{\Span{a}^\perp}^{\tilde\alpha_2} }{4} 
$$ 
where $I_v=I$, $I_w=\tilde \Omega$, $s_v=\gamma(\tilde v)$, $s_w=\tilde\gamma'(\tilde w)$, 
\begin{align}\label{eq:1b}
\begin{split}
\tilde \alpha_0(a) = \tilde \alpha_1(a) =\tilde \alpha_2(a) = s_a \\
\tilde \alpha_0(\tilde v) = \tilde \alpha_1(\tilde v)=\tilde\alpha_2(\tilde v)= \gamma(\tilde v)\\
\tilde \alpha_0(\tilde w) = \tilde \alpha_1(\tilde w)=1+\tilde \alpha_2(\tilde w)= \tilde \gamma'(\tilde w) \\
1+\tilde \alpha_0(\tilde v+\tilde w)=  \tilde\alpha_1(\tilde v+\tilde w) =  \tilde \alpha_2(\tilde v+\tilde w) .
\end{split}
\end{align}

 Case II follows from
$$
\frac{\Pi_{a,s} A_{I,\Omega}^\gamma \Pi_{a,s} }{\Tr(A_{I,\Omega}^\gamma \Pi_{a,s})} = \frac{1}{2} A_{I\times a}^{\gamma\times a} +\frac{1}{4}  \left( \delta_{s_a,\tilde \gamma'(a)} A_{\tilde \Omega}^{\tilde \gamma'} - \delta_{s_a,1+\tilde\gamma'(a)} A_{\tilde \Omega}^{\tilde\gamma''} \right)  
$$
since $\tilde\gamma''(v)=\tilde\gamma'(v)+1$ for $v\in \Omega-\set{0}$. Here $\tilde \Omega=\Span{a}^\perp$. For (2.a)  Lemma \ref{lem:decomposition} part (4) and Eq.~(\ref{eq:vertex-map}) imply that
$$
\frac{\Pi_{a,s} A_{I,\Omega}^\gamma \Pi_{a,s} }{\Tr(A_{I,\Omega}^\gamma \Pi_{a,s})}  = 
\frac{1}{2} A_{I\times a}^{\gamma\times a} - \frac{1}{4} A_{\tilde \Omega}^{\tilde \gamma''} =  A_{\Span{a}^\perp}^{\tilde \alpha'}
$$
where $I_v=I\times a$, $I_w$  is one of the other two isotropics,  
\begin{align}\label{eq:2a}
\begin{split}
\tilde\alpha'(a) = s_a\\
\tilde \alpha'(\tilde v) = \tilde\gamma''(\tilde v) \\
\tilde \alpha'(\tilde w) = 1+\tilde\gamma''(\tilde w)\\
\tilde \alpha'(\tilde v+\tilde w) = 1+\tilde\gamma''(\tilde v+\tilde w).
\end{split}
\end{align}
For (2.a) Lemma \ref{lem:decomposition} part (4) and Eq.~(\ref{eq:vertex-map}) imply that
$$
\frac{\Pi_{a,s} A_{I,\Omega}^\gamma \Pi_{a,s} }{\Tr(A_{I,\Omega}^\gamma \Pi_{a,s})}  = \frac{1}{2} A_{I\times a}^{\gamma\times a} +\frac{1}{4} A_{\tilde \Omega}^{\tilde \gamma'} = \frac{2A_{\Span{a}^\perp}^{\tilde \alpha} + A_{\Span{a}^\perp}^{\tilde\alpha'} }{3} 
$$
where $I_v=I\times a$, $I_w$  is one of the other two isotropics, $\tilde \alpha= \tilde \gamma'$, 
\begin{align}\label{eq:2b}
\begin{split}
\tilde\alpha'(a) = s_a\\
\tilde \alpha'(\tilde v) = \tilde\gamma'(\tilde v) \\
\tilde \alpha'(\tilde w) = 1+\tilde\gamma'(\tilde w)\\
\tilde \alpha'(\tilde v+\tilde w) = 1+\tilde\gamma'(\tilde v+\tilde w).
\end{split}
\end{align}


 Case III follows from the following computation  
$$
\Pi_{a,s} A_{I,\Omega}^\gamma \Pi_{a,s} = \frac{1}{2} A_{I\times a}^{\gamma\times a} +\frac{\delta_{s_a,\tilde \gamma'(a)}}{4}  \left(  A_{\tilde \Omega}^{\tilde \gamma'} -   A_{\tilde \Omega}^{\tilde\gamma''} \right) . 
$$
We have $\tilde \gamma'(a)=\tilde\gamma''(a)$ since $a$  can be written as $a=v+w$ with $v,w\in \Omega-\set{0}$. In this case $I$ and $\tilde\Omega $ are maximal isotropics contained in $\Span{a}^\perp$ intersecting at $\Span{a}$. For (3.a) Lemma \ref{lem:decomposition} part (1) and Eq.~(\ref{eq:vertex-map}) imply that
$$
\frac{\Pi_{a,s} A_{I,\Omega}^\gamma \Pi_{a,s} }{\Tr(A_{I,\Omega}^\gamma \Pi_{a,s})}  =  \frac{1}{2} A_{I\times a}^{\gamma\times a} = \frac{A_{\Span{a}^\perp}^{\tilde \alpha_0}+ A_{\Span{a}^\perp}^{\tilde \alpha_1}  }{2}
$$
where $I_v=I\times a$, $I_w=\tilde \Omega$, $s_v=\gamma(\tilde v)$, 
\begin{align}\label{eq:3a}
\begin{split}
\tilde \alpha_0(a) = \tilde \alpha_1(a) = s_a \\
\tilde \alpha_0(\tilde v) = \tilde \alpha_1(\tilde v)=\gamma(\tilde v)\\
 \tilde \alpha_1(\tilde w) = 1+ \tilde \alpha_0(\tilde w)\\
  \alpha_1(\tilde v+\tilde w) = 1+\tilde \alpha_0(\tilde v+\tilde w)
  \end{split}
\end{align}
For (3.b) Lemma \ref{lem:decomposition} part (2) and Eq.~(\ref{eq:vertex-map}) imply that  
$$
\frac{\Pi_{a,s} A_{I,\Omega}^\gamma \Pi_{a,s} }{\Tr(A_{I,\Omega}^\gamma \Pi_{a,s})}  =   \frac{1}{2} A_{I\times a}^{\gamma\times a} +\frac{1}{4} (A_{\tilde \Omega}^{\tilde \gamma'} - A_{\tilde \Omega}^{\tilde \gamma''} )  =     \frac{A_{\Span{a}^\perp}^{\tilde\alpha_0} +  A_{\Span{a}^\perp}^{\tilde\alpha_1}}{2} 
$$
where $I_v=I\times a$, $I_w=\tilde \Omega$, $s_v=\gamma(\tilde v)$, $s_w=\tilde \gamma'(\tilde w)$,  
\begin{align}\label{eq:3b}
\begin{split}
\tilde \alpha_0(a) = \tilde \alpha_1(a)  = s_a \\
\tilde \alpha_0(\tilde v) = \tilde \alpha_1(\tilde v)=  \gamma(\tilde v)\\
\tilde \alpha_0(\tilde w) = \tilde \alpha_1(\tilde w)=  \tilde \gamma'(\tilde w) \\
1+\tilde \alpha_0(\tilde v+\tilde w) = \tilde \alpha_1(\tilde v+\tilde w) .
\end{split}
\end{align}    

Note that the case where $a\in I$ and $a\in\Omega$ does not occur since $I\cap \Omega=0$. $\Box$
 
 


\begin{thebibliography}{99}

\bibitem{sup}
R. Cleve, A. Ekert, C. Macchiavello and M. Mosca, {\em{Quantum Algorithms Revisited}}, Proc. R. Soc. A \textbf{454}, 339 (1998). 

\bibitem{Vidal1}
G. Vidal, {\em{Efficient Classical Simulation of Slightly Entangled Quantum Computations}}, Phys. Rev. Lett. \textbf{91}, 147902 (2003).

\bibitem{VdN}
M. Van den Nest, W. D{\"u}r, G. Vidal, and H. J. Briegel, {\em{Classical simulation versus universality in measurement-based quantum computation}}, Phys. Rev. A \textbf{75}, 012337  (2007).

\bibitem{Feyn}
R. P. Feynman, {\em{Simulating physics with computers}}, Int. J. Phys. 21, 467 (1982).

\bibitem{Stab} 
D. Gottesman, Proceedings of the XXII International Colloquium on Group Theoretical Methods in Physics,
p. 32-43 (Cambridge, MA, International Press, 1999).

\bibitem{VdN2}
M. Van den Nest, {\em{Universal Quantum Computation with Little Entanglement}}, Phys. Rev. Lett. \textbf{110}, 060504 (2013).

\bibitem{TETBU}
D. Gross, S. T. Flammia, and J. Eisert, {\em{Most Quantum States Are Too Entangled To Be Useful As Computational Resources}}, Phys. Rev. Lett. \textbf{102}, 190501 (2009).

\bibitem{Illu}
D. Poulin, A. Qarry, R. Somma, and F. Verstraete, {\em{Quantum Simulation of Time-Dependent Hamiltonians and the Convenient Illusion of Hilbert Space}}, Phys. Rev. Lett. \textbf{106}, 170501 (2011).


\bibitem{HVM}
M.~Zurel, C.~Okay, R.~Raussendorf, {\em{A hidden variable model for universal quantum computation with magic states on qubits}},
Phys. Rev. Lett. \textbf{125}, 260404 (2020).  
 

\bibitem{loc}
R. Jozsa and N. Linden, {\em{On the Role of Entanglement in Quantum-Computational Speed-Up}},  Proc. R. Soc. Lond. A \textbf{459}, 2011 (2003).

\bibitem{MG}
L.G. Valiant, {\em{Expressiveness of matchgates}},  Theor. Comp. Sci. \textbf{289}, 457 (2002).

\bibitem{JW2}
R. Jozsa, A. Miyake, {\em{Matchgates and classical simulation of quantum circuits}}, Proc. R. Soc. A \textbf{464}, 3089-3106 (2008).

\bibitem{NegWi}
V. Veitch, C. Ferrie, D. Gross, J. Emerson, {\em{Negative Quasi-Probability as a Resource for Quantum Computation}}, New J. Phys. 14, 113011 (2012).


\bibitem{M1}
M. Howard, J. Wallman, V. Veitch and J. Emerson, {\em{Contextuality supplies the `magic' for quantum computation}}, Nature \textbf{510}, 351 (2014).

\bibitem{Hakop}
H. Pashayan, J.J. Wallman, S.D. Bartlett, {\em{Estimating Outcome Probabilities of Quantum Circuits Using Quasiprobabilities}}, Phys. Rev. Lett. \textbf{115}, 070501 (2015).


\bibitem{M1b}
V. Veitch, S.A.H. Mousavian, D. Gottesman and J. Emerson, {\em{The resource theory of stabilizer quantum computation}}, New J. Phys. \textbf{16}, 013009 (2014).

\bibitem{M2}
M. Howard, E.T. Campbell, {\em{Application of a Resource Theory for Magic States to Fault-Tolerant Quantum Computing}}, Phys. Rev. Lett. \textbf{118}, 090501 (2017).

\bibitem{M3}
M. Heinrich, D. Gross, {\em{Robustness of Magic and Symmetries of the Stabiliser Polytope}}, Quantum \textbf{3}, 132 (2019).

\bibitem{M4}
S. Bravyi and D. Gosset, {\em{Improved classical simulation of quantum circuits dominated by Clifford gates}}, Phys. Rev. Lett. \textbf{116} 250501( 2016).


\bibitem{M5}
S. Bravyi, D. Browne, P. Calpin, E. Campbell, D. Gosset, M. Howard, {\em{Simulation of quantum circuits by low-rank stabilizer decompositions}}, Quantum \textbf{3}, 181 (2019).


\bibitem{Galv1}
E.F. Galv{\~a}o, Phys. Rev. A \textbf{71}, 042302 (2005).

\bibitem{Galv2}
C. Cormick, E.F. Galv{\~a}o, D. Gottesman, J.P. Paz, and A.O. Pittenger,  Phys. Rev A \textbf{73}, 012301 (2006).


\bibitem{ReWi} 
N. Delfosse, P. Allard Guerin, J. Bian and R. Raussendorf, Phys. Rev. X \textbf{5}, 021003 (2015).

\bibitem{raussendorf2020phase}
R.~Raussendorf, J.~Bermejo-Vega, E.~Tyhurst, C.~Okay, and M.~Zurel.
\newblock Phase-space-simulation method for quantum computation with magic
  states on qubits.
\newblock {\em Physical Review A}, 101(1):012350, 2020.  
  
\bibitem{kirby2019contextuality}
W.~M.~Kirby, and P.~J.~Love, {\em{Contextuality test of the nonclassicality of variational quantum eigensolvers}}, Phys. Rev. Lett. \textbf{123}, 200501 (2019). 
 

\bibitem{GT}
  D. Gottesman and I.L. Chuang, {\em{Demonstrating the viability of universal quantum computation using teleportation and single-qubit operations}}, Nature \textbf{402}, 390 (1999).
  
\bibitem{NC}
M.A. Nielsen and I.L. Chuang, {\em{Quantum Computation and Quantum Information}}, Cambridge University Press (2000).  
  
 \bibitem{BK}
S. Bravyi and A. Kitaev, Phys. Rev. A \textbf{71}, 022316 (2005).

   
\bibitem{Heim}
A. Heimendahl, MSc thesis, University of Cologne (2019).

\bibitem{PBR}
M.~Pusey, J.~Barrett, T.~Rudolph, Nat.~Phys.~\textbf{8}, 475-478 (2012).
 



\bibitem{Grun13}
B.~Gr{\"u}nbaum.
\newblock {\em Convex polytopes},  {\em Springer Science \& Business Media}.
\newblock volume~221, 2013.

 

\bibitem{BravyiSmolin2016}
S.~Bravyi, G.~Smith, and J.A.~Smolin, Phys.~Rev.~X \textbf{6}, 021043 (2016).




\end{thebibliography}
\end{document}